\documentclass[aps,prb,amsmath,amssymb,nofootinbib,twocolumn,
superscriptaddress]{revtex4-1}

\usepackage{color,array}
\usepackage{slashed}
\usepackage{graphicx}
\usepackage{epsfig}
\usepackage{bm}
\usepackage{psfrag}
\usepackage{hyperref}
\graphicspath{{../fig/}}
\newcommand{\be}{\begin{equation}}
\newcommand{\ee}{\end{equation}}
\newcommand{\ben}{\begin{eqnarray}}
\newcommand{\een}{\end{eqnarray}}

\newcommand{\ket}[1]{| #1\rangle}

\newcommand{\vect}[1]{\boldsymbol{#1}}

\begin{document}

\title{Spin dynamics of the bilinear-biquadratic $S=1$ Heisenberg model on the triangular lattice: a quantum Monte Carlo study}
\date{\today}

\author{Annika V\"oll}
\affiliation{\mbox{\it Institut f\"ur Theoretische Festk\"orperphysik, JARA-FIT and JARA-HPC, RWTH Aachen University, 52056 Aachen, Germany}}

\author{Stefan Wessel}
\affiliation{\mbox{\it Institut f\"ur Theoretische Festk\"orperphysik, JARA-FIT and JARA-HPC, RWTH Aachen University, 52056 Aachen, Germany}}

\begin{abstract}
We study thermodynamic properties as well as the dynamical spin and quadrupolar structure factors  of the O(3)-symmetric spin-1 Heisenberg model with bilinear-biquadratic exchange interactions on the triangular lattice. Based on a sign-problem-free quantum Monte Carlo approach, we access both the ferromagnetic and the ferroquadrupolar ordered, spin nematic phase as well as the 
SU(3)-symmetric point which separates these phases.
Signatures of Goldstone soft-modes in the dynamical spin and the quadrupolar structure factors are identified, and the properties of the low-energy excitations 
are compared to the thermodynamic behavior observed at finite temperatures as well as to Schwinger-boson flavor-wave theory. 
\end{abstract}

\maketitle

\section{Introduction}
In recent years, the emergence of unconventional order in magnetic systems has been investigated  both experimentally as well as from various theoretical perspectives. In contrast to quantum spin liquids~\cite{balents10}, which are characterized by the absence of any symmetry-breaking long-range order, spin nematic states exhibit long-ranged quadrupolar correlations without conventional (dipolar) magnetic order, and thus also do not feature magnetic Bragg peaks~\cite{andreev84, chubukov90}. While time-reversal symmetry remains intact, the spontaneous breaking of the spin rotation symmetry in spin nematic phases leads to the emergence of low-energy Goldstone modes with a linear dispersion relation, providing an algebraic contribution to the low-temperature ($T$) specific heat. 
Spin nematic phases have indeed been identifies in several model systems, such as in spin-1/2 Heisenberg systems with competing ferromagnetic and antiferromagnetic or ring-exchange interactions~\cite{laeuchli05, shannon06}, as well as in spin-1 Heisenberg models with sizable biquadratic interaction terms~\cite{mila00} in addition to the bilinear spin exchange interaction~\cite{harada02,bhattacharjee06,laeuchli06,tsunetsugu06}.

One particular motivation for various recent studies of the triangular lattice spin-1 Heisenberg model with additional biquadratic exchange terms  was its initially proposed relevance to  the unconventional magnetic properties observed in the Ni-based compound NiGa${}_2$S${}_4$~\cite{nakatsuji05}. In this material, spin-1 carrying Ni${}^{2+}$-ions reside on weakly coupled triangular lattice layers. While no signatures of low-temperature dipolar magnetic long-range order was detected,  NiGa${}_2$S${}_4$ exhibits a $T^2$-dependent low-temperature specific heat, akin to linearly dispersing two-dimensional low-energy soft modes~\cite{nakatsuji05}. 
Within  the spin-1 bilinear-biquadratic Heisenberg model scenario, two distinct spin nematic states have been identified and were proposed as possible ground states for NiGa${}_2$S${}_4$. These states differ in the relative orientation of the local director, specifying the axis perpendicular to the plane of dominant residual local spin fluctuations. While the ferroquadrupolar state~\cite{bhattacharjee06,laeuchli06} is characterized by a uniform alignment of the directors, akin to the uni-axial nematic liquid crystal state, the antiferroquadrupolar state~\cite{tsunetsugu06} exhibits a three-sublattice structure with  mutually perpendicular directors on neighboring lattice sites. 
A variety of experimental techniques, applied to NiGa${}_2$S${}_4$, however revealed a dynamical freezing with incommensurate short-range order of local magnetic moments that form at low temperatures, and alternative scenarios have been put forward to explain the peculiar properties of this compound (cf. Ref.~\onlinecite{nakatsuji10} for an overview).
Even in the absence of true long-range quadrupolar order, the emergence of substantial quadrupolar correlations may still be employed to rationalize e.g. the two-peak structure observed in the specific heat of NiGa${}_2$S${}_4$~\cite{stoudenmire09}.

It therefore appears relevant, also from a more general perspective, to provide further theoretical characterizations of quadrupolar order for future probes based, e.g., on inelastic neutron or light scattering experiments, nuclear magnetic resonance techniques, and possibly resonant X-ray scattering~\cite{michaud11,smerald13,smerald13b}.  
Several studies of the dynamical spin and quadrupolar structure factors in triangular-lattice-based spin nematic phases have indeed been performed, based on approximate schemes to calculate these dynamical quantities, e.g. within Schwinger-boson flavor-wave theory~\cite{papanicolau84,joshi99,tsunetsugu06,laeuchli06,pires14}. The quality of such approximate schemes may be assessed by employing alternative methods, applicable within appropriate model parameter regimes. 
While for an underlying square lattice geometry, the spin-1 bilinear-biquadratic Heisenberg model and its thermodynamic properties have been studied by unbiased, large-scale quantum Monte Carlo (QMC) simulations over a wide range of model parameters~\cite{kawashima04,harada01,harada02,harada07}, the non-bipartiteness of the triangular lattice typically leads to a severe QMC sign-problem for Heisenberg-like spin models. Only recently was a sign-problem-free QMC representation of the partition function formulated for the pure biquadratic exchange model~\cite{kaul12}.

Here, we expand on the approach of Ref.~\onlinecite{kaul12}, and cover an extended region of parameter space of the spin-1 bilinear-biquadratic Heisenberg model on the triangular lattice, which covers both the ferroquadrupolar and the ferromagnetically ordered region as well as the SU(3)-symmetric point that separates both phases. 
Beyond an identification of the ground state order parameters and finite temperature thermodynamic properties, we in particular employ the QMC algorithm to measure the dynamical structure factors for both spin correlations as well as quadrupolar correlation functions. This allows us to identify features of these phases in the dynamical spin structure factors related to, e.g., neutron scattering experiments, as well as to quantitatively relate these spectral properties to the thermodynamic behavior. We furthermore compare our numerical findings to the results from flavor-wave theory~\cite{laeuchli06}, which in addition provides us with exact results at the SU(3) point. 

The remainder of this article is organized as follows: In Sec. II, we present details on the considered model and the employed numerical method. Then, in Sec. III, we present our QMC results on the order parameters and the thermodynamic properties. The dynamical structure factors are presented in Sec. III, and analyzed in terms of the low-energy modes and the corresponding thermodynamic properties, with some final conclusions given in Sec. V.

\section{Model and Method}
We consider in the following the spin-$1$ bilinear-biquadratic Heisenberg model on the triangular lattice, described in terms of localized spin-1 degrees of freedom $\vect{S}_i$ on each lattice site $i$ by the Hamiltonian
\be\label{hamiltonian}
H=J\sum_{\langle i,j \rangle} \left[\cos\theta\: \vect{S}_i\cdot\vect{S}_j + \sin\theta\left(\vect{S}_i\cdot\vect{S}_j\right)^2 \right],
\ee
parametrized by the angular parameter $\theta$, following conventional notations~\cite{laeuchli06}. In the following, we set $J=1$, and concentrate on the parameter regime $\theta\in[-\pi,-\pi/2]$, which is accessible by sign-problem-free QMC simulations, as detailed below. This parameter range includes the SU(3)-symmetric point, $\theta=\theta_{SU(3)}=-3\pi/4$, that separates (within the considered parameter range) a ferromagnetic phase for $\theta\in[-\pi,\theta_{SU(3)}]$ from a spin nematic phase with pure ferroquadrupolar order for $\theta\in(\theta_{SU(3)},\pi/2)$ (see e.g. Ref.~\onlinecite{laeuchli06} for a detailed study of the full quantum phase diagram of the Hamiltonian $H$ and the full extent of both phases).

Our QMC algorithm is based on the directed loop stochastic series expansion approach~\cite{syljuasen02}. In order to obtain a sign-problem-free representation of the quantum partition function $Z=\mathrm{Tr}\exp(-\beta H)$, with the inverse temperature $\beta=1/T$, we require all matrix elements of $(-H)$ to be non-negative. In the parameter region of the ferromagnetic phase,  $\theta\in[-\pi,\theta_{SU(3)}]$, this is feasible in the standard local-$S^z$ basis with eigenstates $\ket{1},\ket{0},\ket{\bar{1}}$ of $S^z$ for the eigenvalues $1,0,-1$, respectively. In this parameter region, all off-diagonal matrix elements of $H$ are non-positive, and we merely require to subtract a constant $C\geq -\cos\theta+2\sin\theta$ from $H$ in order to ensure the shifted $-H'=-H+C$ to be non-negative.
For $\theta\in(\theta_{SU(3)},\pi/2]$, the Hamiltonian $H$ exhibits positive off-diagonal matrix elements. In order to avoid the sign-problem in this parameter region on the non-bipartite triangular lattice, we employ the idea of Ref.~\onlinecite{kaul12}, and consider the phase-rotated basis state $\ket{0'}=i\ket{0}$. Furthermore, we subtract a constant $C\geq \sin\theta$ to ensure the shifted Hamiltonian $H'$ to be non-positive in the rotated basis (this basis rotation requires to be accounted for when measuring observables). It is worth noticing that for $\theta=-\pi/2$ the matrix elements of the shifted Hamiltonian $H'$ are either $0$ or $-1$ in the rotated basis. In fact, the two-site bond Hamiltonian for $\theta=-\pi/2$ equals a projector,
\[
-H'|_{\theta=-\pi/2}=\left( \ket{0'0'}+\ket{1\bar{1}}+\ket{\bar{1}1} \right)\left( \ket{0'0'}+\ket{1\bar{1}}+\ket{\bar{1}1}\right),
\] 
and the directed loop equations allow for a QMC sampling in terms of a three-color closed-loop representation~\cite{kaul12}.

In our simulations, we considered finite systems with $N=L^2$ lattice sites and periodic boundary conditions in both lattice directions $\vect{a}_1=(a,0)^\intercal$, and $\vect{a}_2=(a/2,\sqrt{3}a/2)^\intercal$, where the lattice constant is set to $a=1$ in the following.
In order to access ground state properties, $\beta$ must  be chosen sufficiently large. We typically require $\beta\geq L$ for this purpose. 
Besides the energy $E=\langle H \rangle$ and the specific heat $C_V=(dE/dT)/N$, we measured the uniform susceptibility,
\be
\chi_\mathrm{u}=\frac{\beta}{N}\left\langle \left( \sum_{i=0}^N S^z_i \right)^2 \right\rangle,
\ee
as well as the (equal-time) spin structure factor
\be\label{ssq}
S_S(\vect{q})=\frac{1}{3}\frac{1}{N}\sum_{i,j} e^{-i\vect{q}\cdot(\vect{r}_i-\vect{r}_j)} \left\langle \vect{S}_i\cdot \vect{S}_j \right\rangle,
\ee
with $\vect{r}_i$ denoting the position of the $i$-th lattice site. A prefactor $1/3$ was inserted in Eq.~(\ref{ssq}) to comply with the 
notation of Ref.~\onlinecite{kaul12}. Within the QMC simulations, we employ the O(3) invariance of $H$, to access 
$S_S(\vect{q})$ through the diagonal operator components ${S}^{z}_i$ as
\be
S_S(\vect{q})=\frac{1}{N}\sum_{i,j} e^{-i\vect{q}\cdot(\vect{r}_i-\vect{r}_j)} \left\langle S^z_i S^z_j \right\rangle.
\ee
The spin structure factor allows to detect the presence of (dipolar) magnetic order, indicated in the ferromagnetic phase by an extensive scaling of $S_S(\vect{q})$ at the ferromagnetic Bragg peak position $\vect{q}=0$. 
In order to access quadrupolar order, we also consider the correlation function $\langle \sum_{\alpha\beta} Q_i^{\alpha\beta} Q_j^{\alpha,\beta}\rangle=\langle \mathrm{Tr} [{Q}_i {Q}_j]\rangle$ of the magnetic quadrupolar moments, given by the traceless, symmetric $3\times 3$  matrix operator $Q_i$ with elements
\be
Q_i^{\alpha\beta}=\frac{1}{2}\left(S_i^\alpha S_i^\beta + S_i^\beta S_i^\alpha \right)-\frac{2}{3}\delta_{\alpha\beta},
\ee
with $\delta_{\alpha\beta}$ the Kronecker delta, and $\alpha,\beta \in \{x,y,z\}$. Due to the explicit O(3) symmetry of the Hamiltonian $H$, we can probe for the  O(3)-symmetric quadrupolar correlations upon measuring the trace of the quadrupolar structure factor
\be\label{sqq}
S_Q(\vect{q})=\frac{2}{5}\frac{1}{N}\sum_{i,j} e^{-i\vect{q}\cdot(\vect{r}_i-\vect{r}_j)} \left\langle \mathrm{Tr} [{Q}_i {Q}_j] \right\rangle
\ee
in the QMC simulations through the (in the computational basis) diagonal operator components ${Q}^{zz}_i$, as
\be
S_Q(\vect{q})=\frac{3}{N}\sum_{i,j} e^{-i\vect{q}\cdot(\vect{r}_i-\vect{r}_j)} \left\langle {Q}^{zz}_i {Q}^{zz}_j \right\rangle,
\ee
which allows for an efficient implementation. The prefactor $2/5$ in the definition of $S_Q(\vect{q})$ in Eq.~(\ref{sqq}) was chosen in order to comply with the notation of Ref.~\onlinecite{kaul12}. The quadrupolar structure factor allows to detect the presence of quadrupolar order, which in the ferroquadrupolar case is signaled by an extensive scaling of $S_Q(\vect{q})$ at $\vect{q}=0$. Within the ferromagnetic phase, $S_Q(\vect{q})$ also exhibits extensive scaling  at $\vect{q}=0$, with quadrupolar long-ranged correlations induced by the ferromagnetic ordered ground state, whereas $S_S(\vect{q})$ does not exhibit Bragg peaks in the ferroquadrupolar phase. 

\begin{figure}[t]
  \centering
  \includegraphics[width=\columnwidth]{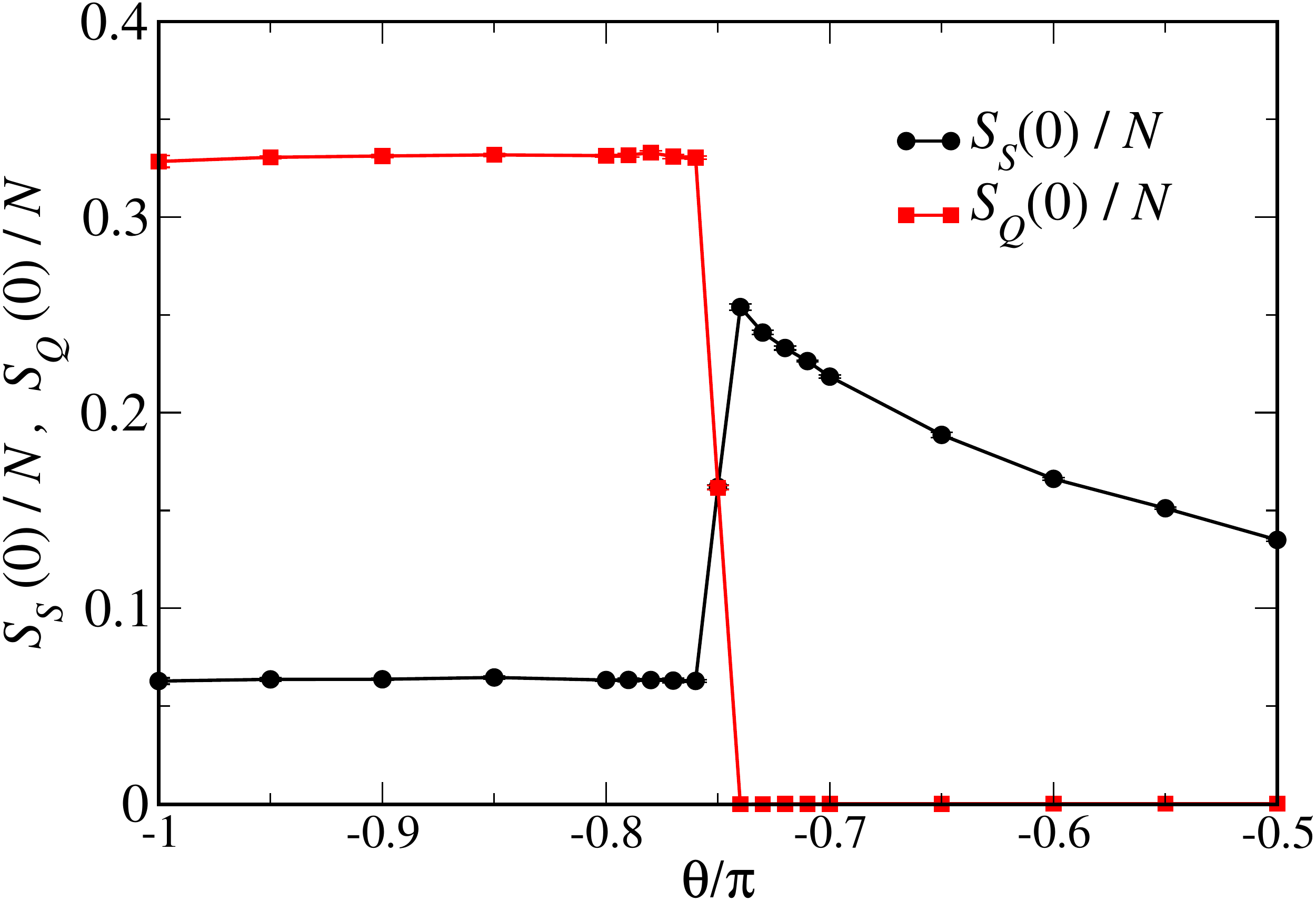}
  \caption{(Color online) Thermodynamic limit results of the $\vect{q}=0$ values of the spin and quadrupolar structure factors $S_S(\vect{q}=0)/N$ and $S_Q(\vect{q}=0)/N$ as functions of $\theta$ within the considered parameter range. }
  \label{fig:ops}
\end{figure}

In order to access information on the excitations in these different regimes, we also extracted from  the QMC simulations
the dynamical spin and quadrupolar structure factors, which are defined through
\be
S_S(\omega,\vect{q})= \frac{1}{N}\int dt\sum_{i,j} e^{i(\omega t - \vect{q}\cdot(\vect{r}_i-\vect{r}_j))} \left\langle {S}_i^{z}(t) {S}^{z}_j(0) \right\rangle,
\ee
and
\be
S_Q(\omega,\vect{q})= \frac{3}{N}\int dt\sum_{i,j} e^{i(\omega t - \vect{q}\cdot(\vect{r}_i-\vect{r}_j))} \left\langle {Q}_i^{zz}(t) {Q}^{zz}_j(0) \right\rangle,
\ee
respectively. Within the QMC simulations, we can efficiently~\cite{michel07} measure the imaginary-time displaced correlation functions directly in Matsubara frequency representation, related via
\be\label{kernel}
S_X(i\omega_n,\vect{q})=\int_0^\infty {d\omega}\: \frac{\omega}{\pi} \frac{(1-e^{-\beta\omega})}{\omega^2+\omega_n^2}\:  S_X(\omega,\vect{q}),
\ee
to the dynamical structure factors for $X=S$ and $Q$, respectively. Here, $\omega_n= 2\pi n /\beta$ for $n=0,1,2,...$ denote the Matsubara frequencies, where typically we require values of $n$ up to 160 to access the leading $1/\omega_n^{2}$ asymptotic behavior of the $ S_X(i\omega_n,\vect{q})$. The numerical inversion of Eq.~(\ref{kernel}) to obtain $S_X(\omega,\vect{q})$ from the Matsubara frequency QMC data $S_X(i\omega_n,\vect{q})$ was performed using the stochastic analytic continuation method~\cite{beach04}. 

\section{Thermodynamic Properties}

\begin{figure}[t]
  \centering
  \includegraphics[width=\columnwidth]{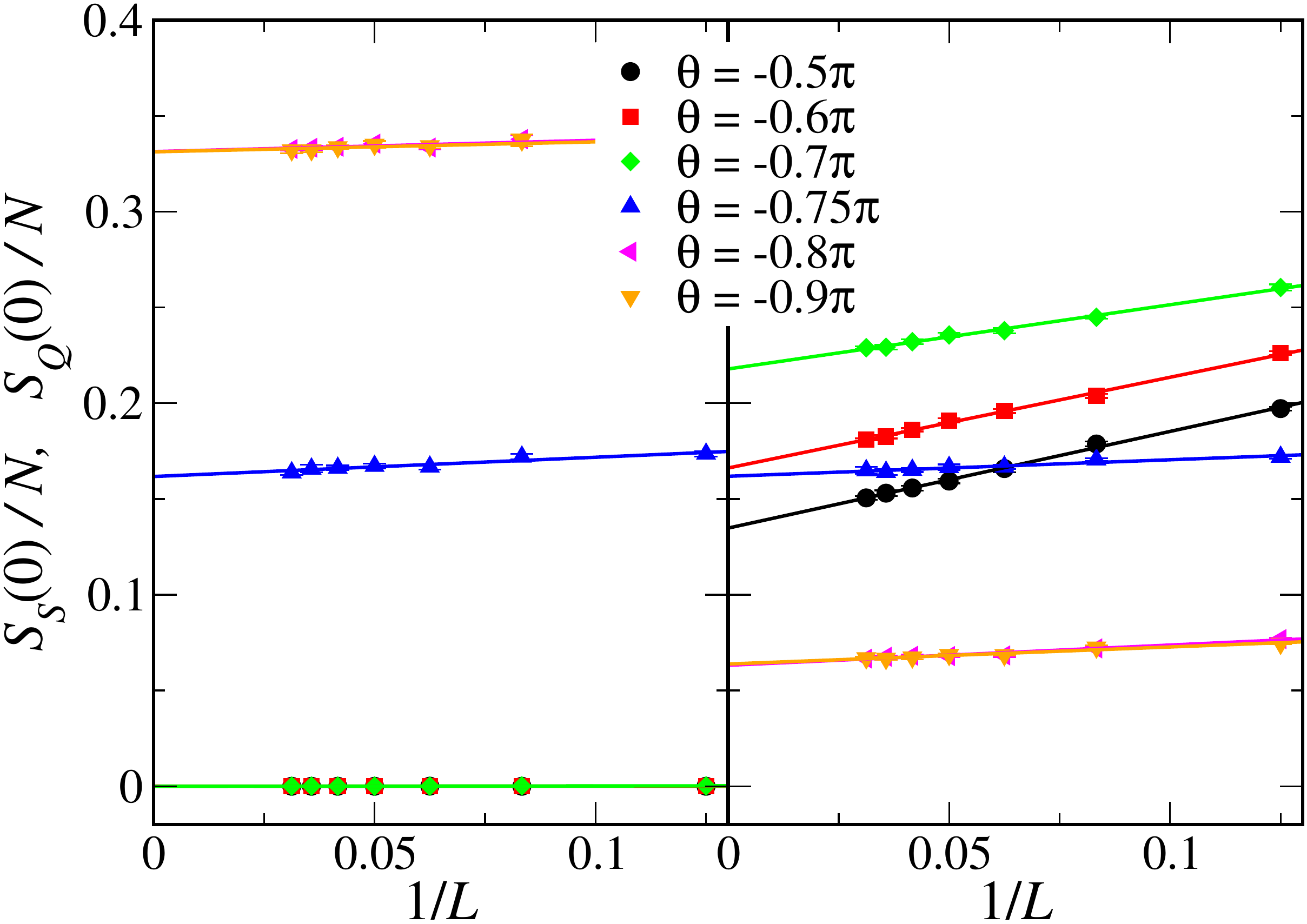}
  \caption{(Color online) Finite size extrapolations of the $\vect{q}=0$ values of the spin and quadrupolar structure factors $S_S(\vect{q}=0)/N$ and $S_Q(\vect{q}=0)/N$ as functions of $1/L$ for various values of $\theta$.}
  \label{fig:opsfs} 
\end{figure}

Before considering these dynamical properties, we establish first the nature of the ground state of the bilinear-biquadratic spin-1 Heisenberg model, Eq.~(\ref{hamiltonian}) on the triangular lattice within the considered parameter region $\theta\in[-\pi,-\pi/2]$. For this purpose, we show in Fig.~\ref{fig:ops} the values of the spin and quadrupolar structure factors at $\vect{q}=0$, obtained from  linear extrapolations in $1/L$ of the finite system values to the thermodynamic limit. For several values of $\theta$, the finite system data and the extrapolations are shown in Fig.~\ref{fig:opsfs}. For $\theta < \theta_{SU(3)}$, a fully polarized ferromagnetic state is exhibited by a saturated value $S_S(\vect{q}=0)/N=1/3$ of the spin structure factor, which leads in effect also to a finite value for the ferroquadrupolar order parameter, indicated by the finite value of $S_Q(\vect{q}=0)/N=1/15$ in this parameters range. At the $SU(3)$-symmetric point, $\theta=\theta_{SU(3)}$,  both structure factors become degenerate, with $S_S(\vect{q}=0)/N=S_Q(\vect{q}=0)/N=1/6$, while for $\theta>\theta_{SU(3)}$, the spin structure factor $S_S(\vect{q}=0)/N$ vanishes (we also verified that no dipolar order at any finite $\vect{q}\neq 0$ appears within $S_S(\vect{q})/N$), whereas $S_Q(\vect{q}=0)/N$ takes on a finite value, characterizing the genuine ferroquadrupolar order in this regime. Our data indicates that $S_Q(\vect{q}=0)/N\rightarrow 4/15$, i.e., the maximum possible value of $S_Q(\vect{q}=0)/N$ for a product state $\prod_i\ket{S^z_i=0}$, in the limit $\theta\rightarrow \theta_{SU(3)}^+$, while quantum fluctuations lead to a reduction of about 50\% from that value for the pure biquadratic model at $\theta=-\pi/2$, in accord with the findings of Ref.~\onlinecite{kaul12}.

\begin{figure}[t]
  \centering
  \includegraphics[width=\columnwidth]{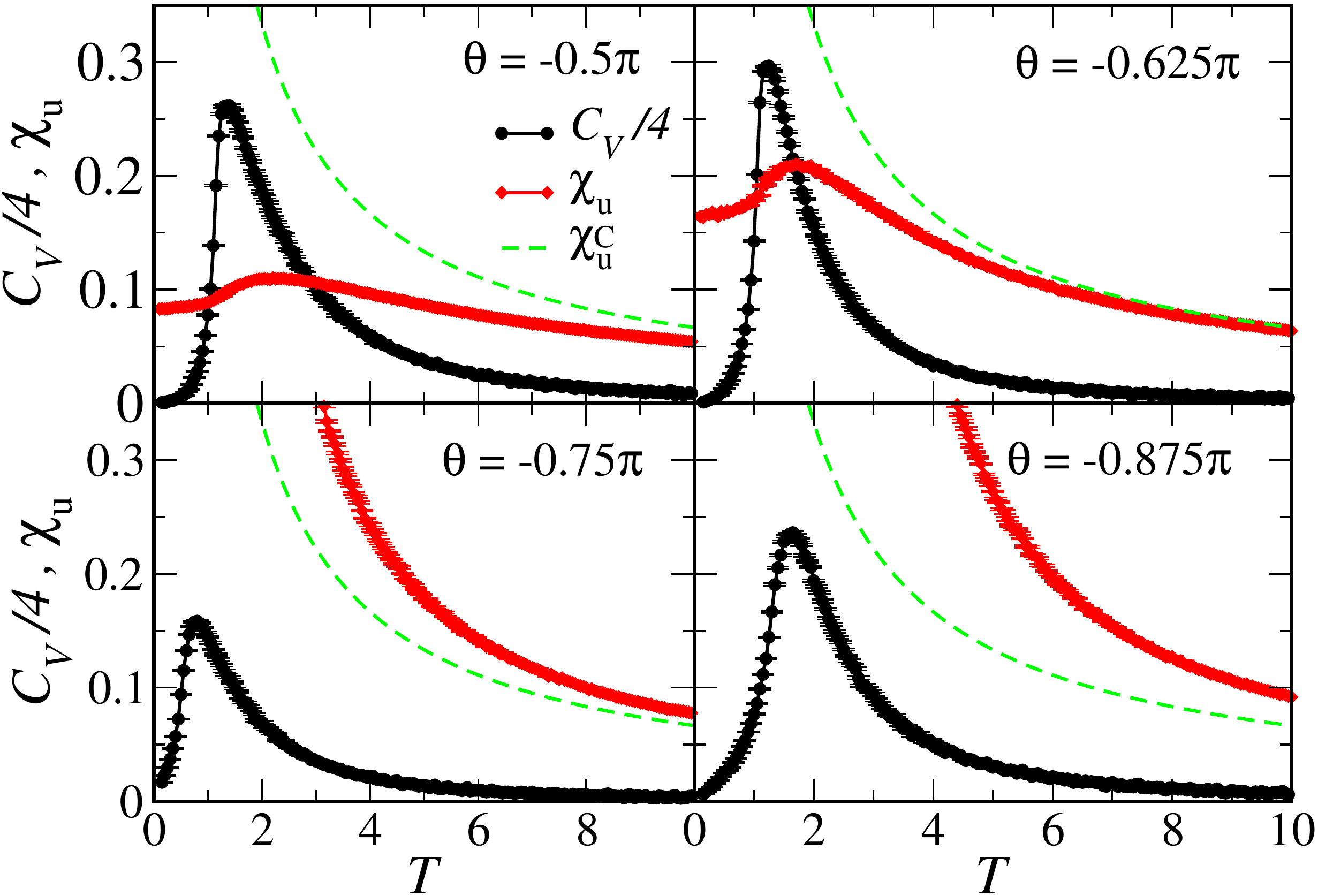}
  \caption{(Color online) Uniform susceptibility $\chi_\mathrm{u}$ and specific heat $C_V$ as functions of temperature $T$ taken for different  values of $\theta$, as indicated. The dashed lines show the $S=1$ single-spin Curie behavior $\chi_\mathrm{u}^\mathrm{C}=S(S+1)/(3T)$ of the uniform susceptibility.}
  \label{fig:chiucv} 
\end{figure}

While at finite temperature the O(3) symmetry is restored, in accord with the Mermin-Wagner theorem~\cite{mermin66}, one may still access characteristic features of the ferromagnetic and ferroquadrupolar phases also from finite-$T$ thermodynamic quantities. For this purpose, we consider in particular the uniform susceptibility $\chi_\mathrm{u}$ and the specific heat $C_V$, shown in Fig.~\ref{fig:chiucv} for different values of $\theta$ as functions of  temperature for an $L=64$ system, representative of the thermodynamic limit behavior in the considered temperature regime.
When comparing our data for the pure biquadratic case, $\theta=-\pi/2$, to the one presented in Ref.~\onlinecite{kaul12}, one notices a deviation in e.g. the Shottky-like peak position of $C_V$ by a factor 3 as compared to the data shown in Fig.~1 of Ref.~\onlinecite{kaul12}. In fact, the data in Ref.~\onlinecite{kaul12} applies to $J=1/3$ instead of the anticipated value of $J=1$~\cite{Kaulpc}. The susceptibility data in Fig.~\ref{fig:chiucv} clearly exhibits the divergent low-$T$ response in the ferromagnetically ordered region, including the SU(3) point,  while $\chi_\mathrm{u}$ saturates to a finite value in the ferroquadrupolar ground state, indicative of the gapless nature of this regime. The comparison to the single-spin $S=1$ Curie behavior in Fig.~\ref{fig:chiucv}, which is approached by the QMC data in the large-$T$ limit, similarly exhibits an enhanced (suppressed) polarizability of the systems within the ferromagnetic (ferroquadrupolar) region. 

\begin{figure}[t]
  \centering
  \includegraphics[width=\columnwidth]{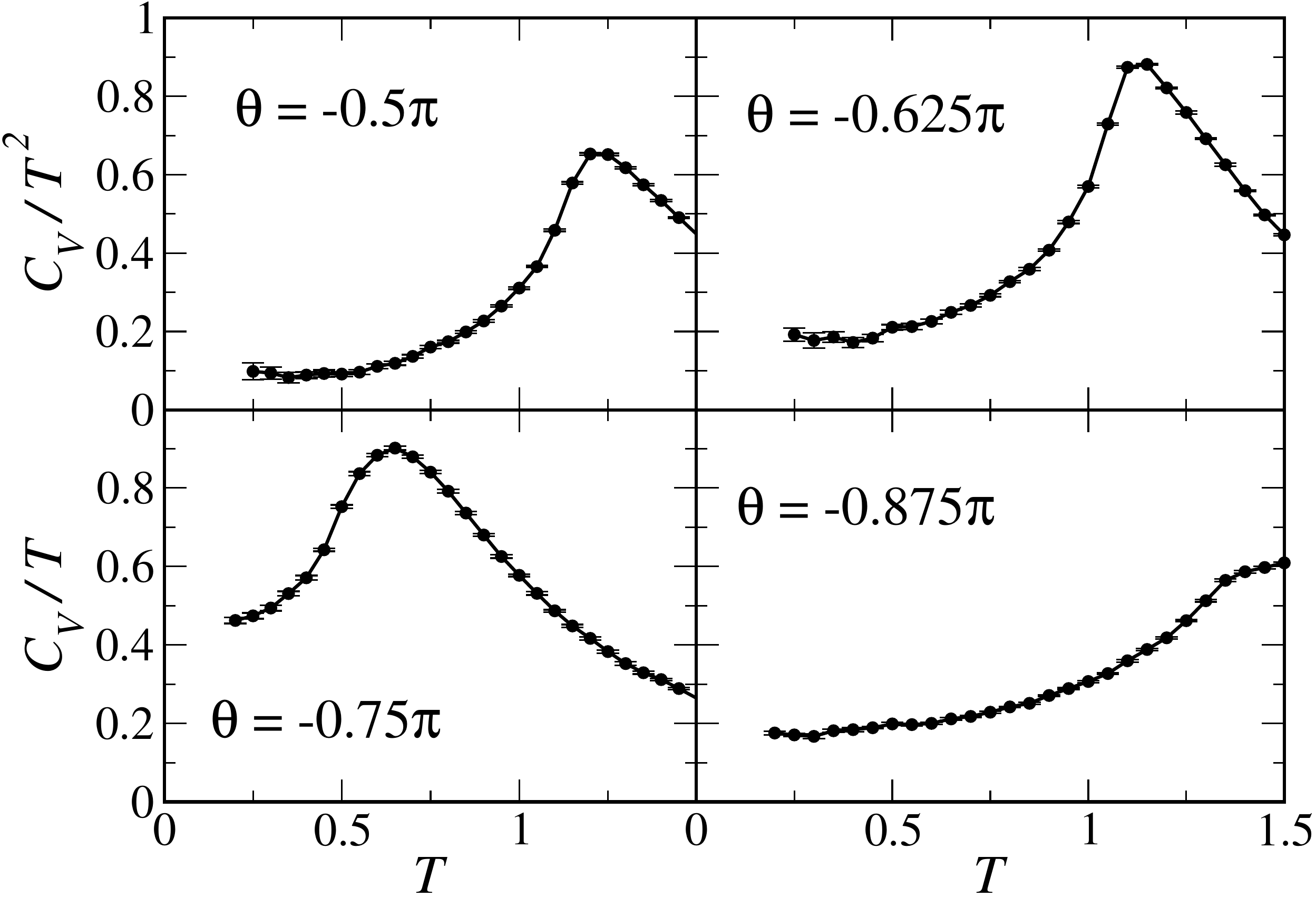}
  \caption{(Color online) Rescaled specific heat $C_V/T$ ($C_V/T^2$) for the ferromagnetic (ferroquadrupolar) region as  a function of temperature $T$ for different  values of $\theta$, as indicated.}
  \label{fig:cvresc} 
\end{figure}

In addition to the Shottky-like peak at $T=O(J)$, the specific heat exhibits for the ferroquadrupolar region  a super-linear suppression at low temperatures, while $C_V$  vanishes linearly with temperature within the ferromagnetic region. In fact, from the dispersions of the low-energy Goldstone modes, considered in more detail in Sec.~IV, we expect the low-$T$ specific heat to exhibit a linear (quadratic) asymptotic low-$T$ scaling within the ferromagnetic (ferroquadrupolar) regime, which can be observed more directly from the appropriately rescaled low-$T$ specific heat data shown in Fig.~\ref{fig:cvresc}. In all considered cases, do the rescaled quantities approach to constant low-$T$ values. 
Anticipating the further discussion in Sec.~IV, we  consider first the pure biquadratic model, $\theta=-\pi/2$, in more detail. For  a single, linearly dispersing low-energy mode, with 
$\omega(\vect{q}) = c |\vect{q}|$, we obtain a leading contribution to the low-$T$ specific heat $C_V=3\:\zeta(3)/\pi \times T^2/c^2$, where $\zeta(3)\approx 1.202$. Employing as an estimate for the  velocity a value of $c\approx 4.8$ (cf. the discussion in Sec.~IV), we obtain a value of $C_V/T^2\approx  0.05$. The data in Fig.~\ref{fig:cvresc} is then in accord with the presence of \textit{two} independent such linear soft-mode contributions. 
Similarly, for $\theta=-0.625\pi$, we obtain $C_V/T^2\approx 0.2$ for two linear soft-modes with $c\approx 3.4$, again in accord with the specific heat data in Fig.~\ref{fig:cvresc}. For $\theta\leq\theta_\mathrm{SU(3)}$,  
the system exhibits quadratic low-energy soft modes (cf. the discussion in Sec.~IV). A single quadratically dispersing low-energy mode with $\omega(\vect{q}) = b \vect{q}^2$ provides
a contribution $C_V=\pi/12\times T/b$ to the low-$T$ specific heat. From the exact result of the low-energy dispersion~\cite{laeuchli06} at the SU(3) point, with $b=3/\sqrt{8}\approx 1.06$ (cf. the discussion in Sec.~IV), we obtain a low-$T$ contribution of $C_V/T\approx0.24$, and the data in Fig.~\ref{fig:cvresc} exhibits the presence of two such quadratic soft-modes in the excitation spectrum at the SU(3) point.  For $\theta=-0.875\pi$, within the ferromagnetic region, we estimate from the data discussed in Sec.~IV a value of $b\approx 1.4$, which leads to a low-$T$ contribution of $C_V/T\approx0.19$. The data in Fig.~\ref{fig:cvresc} is then in accord with the presence of a single  such quadratic soft-mode, reflecting the ferromagnetic ground state for this values of $\theta$.

\section{Dynamical Structure Factors}

\begin{figure}[t]
  \centering
  \includegraphics[width=\columnwidth]{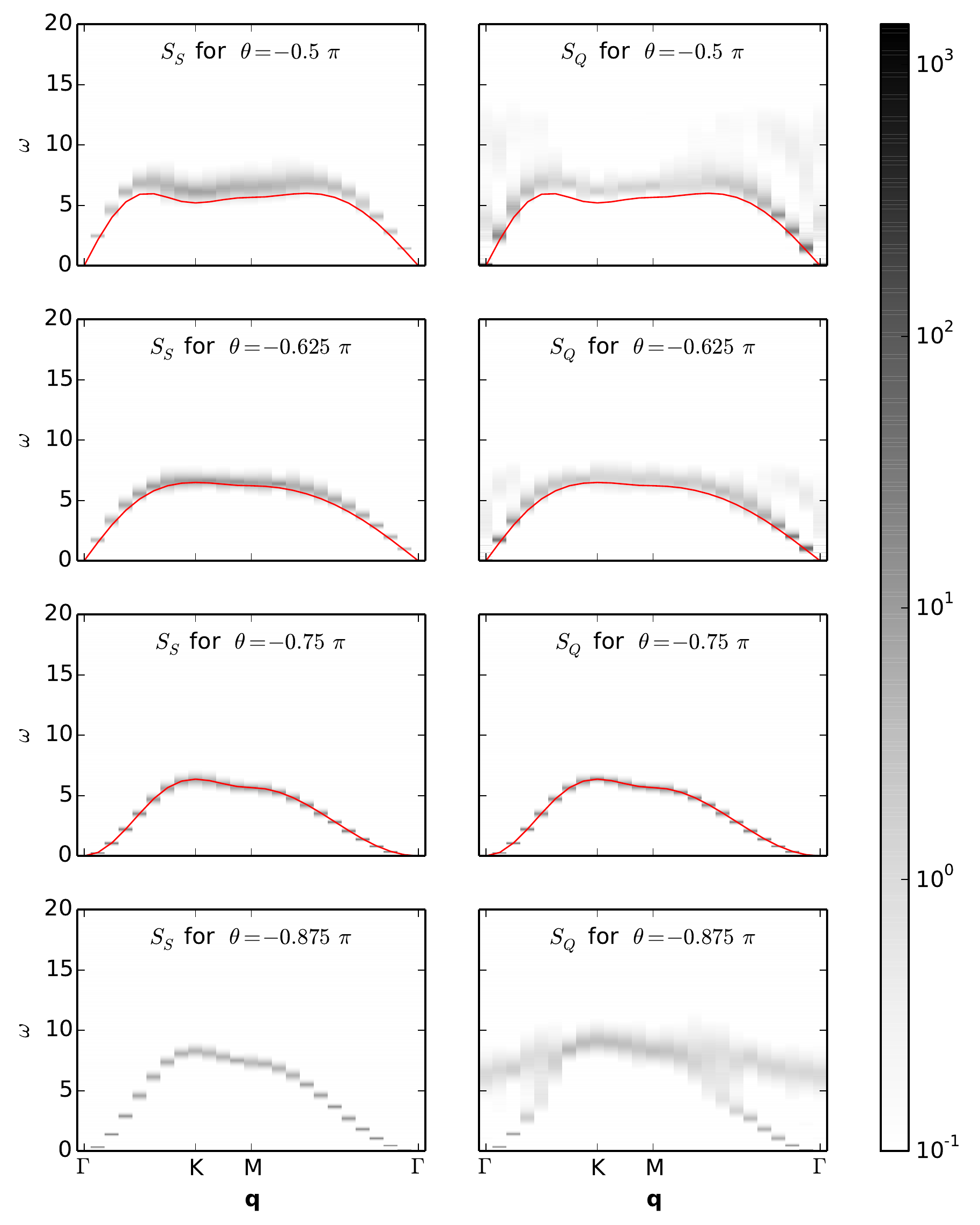}
  \caption{(Color online) Dynamical spin and quadrupolar structure factors $S_S(\omega,\vect{q})$ and $S_Q(\omega,\vect{q})$ for different values of $\theta$ along the path $\Gamma\rightarrow K\rightarrow M \rightarrow \Gamma$ through the Brillouin zone ($\Gamma=(0,0)^\intercal$, $K=(4\pi/3,0)^\intercal$, $M=(2\pi/3,\pi/\sqrt{3})^\intercal$). The dashed lines indicate the quadrupolar wave dispersion relations obtained within flavor-wave theory.}
  \label{fig:dsf} 
\end{figure}

In light of the results in the previous section, we next present our results for the dynamical structure factors. For this purpose, we collect in Fig.~\ref{fig:dsf} both $S_S(\omega,\vect{q})$ and $S_Q(\omega,\vect{q})$ along a standard symmetry path through the triangular lattice Brillouin zone for different values of $\theta$. 
We included in 
Fig.~\ref{fig:dsf} in addition the results for the quadrupolar wave dispersions obtained from the Schwinger-boson flavor-wave theory from Ref.~\onlinecite{laeuchli06}. Within this approach, one finds two degenerate branches of quadrupolar waves, with dispersion relation
$\omega(\vect{q})=\sqrt{A^2(\vect{q})-B^2(\vect{q})}$, where $A(\vect{q})=6 J(\cos\theta \:\gamma({\vect{q}})-\sin\theta)$, $B(\vect{q})=6 J(\sin\theta -\cos\theta)\gamma({\vect{q}})$ and $\gamma({\vect{q}})=\sum_j \exp(i\vect{\delta}_j\cdot\vect{q})/6$, with $\vect{\delta_j}$, $j=1,...,6$ the six nearest neighbor vectors on the triangular lattice. Here, a  factor of 2 missing in  the formula for $A(\vect{q})$ and $B(\vect{q})$ in Ref.~\onlinecite{laeuchli06} has been corrected for~\cite{Pencpc}. From Fig.~\ref{fig:dsf}, we find that our QMC data at the SU(3)-symmetric point fall perfectly onto the flavor-wave theory result, which provides a useful quality check of our numerical procedure. Moving away from the SU(3) point into the quadrupolar phase, flavor-wave theory provides only approximate dispersions. Nevertheless, 
we find that flavor-wave theory still accounts for the  dispersion relation from QMC rather accurately, with a largest deviation in excitation energy of about 15\% for the pure biquadratic model. 

\begin{figure}[t]
  \centering
  \includegraphics[width=\columnwidth]{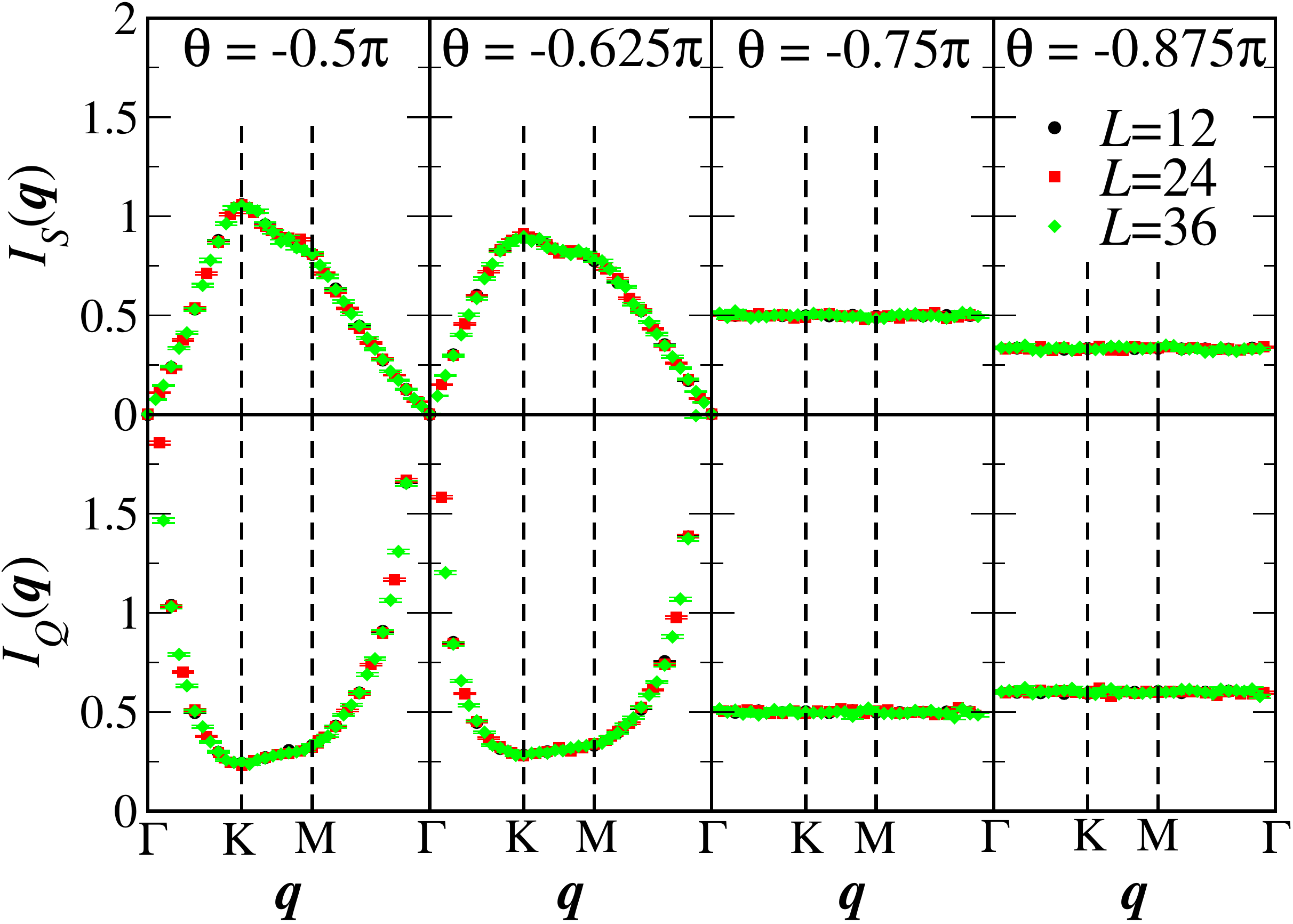}
  \caption{(Color online) Integrated spectral weights $I_S(\vect{q})$ and $I_Q(\vect{q})$ for different values of $\theta$ along the same Brillouin-zone path as in Fig.~5. Values of $I_S(\vect{q})$ and $I_Q(\vect{q})$ at $\vect{q}=\Gamma$  that extend beyond the range of the figure are not on display.}
  \label{fig:sfq} 
\end{figure}
\begin{figure}[t]
  \centering
  \includegraphics[width=\columnwidth]{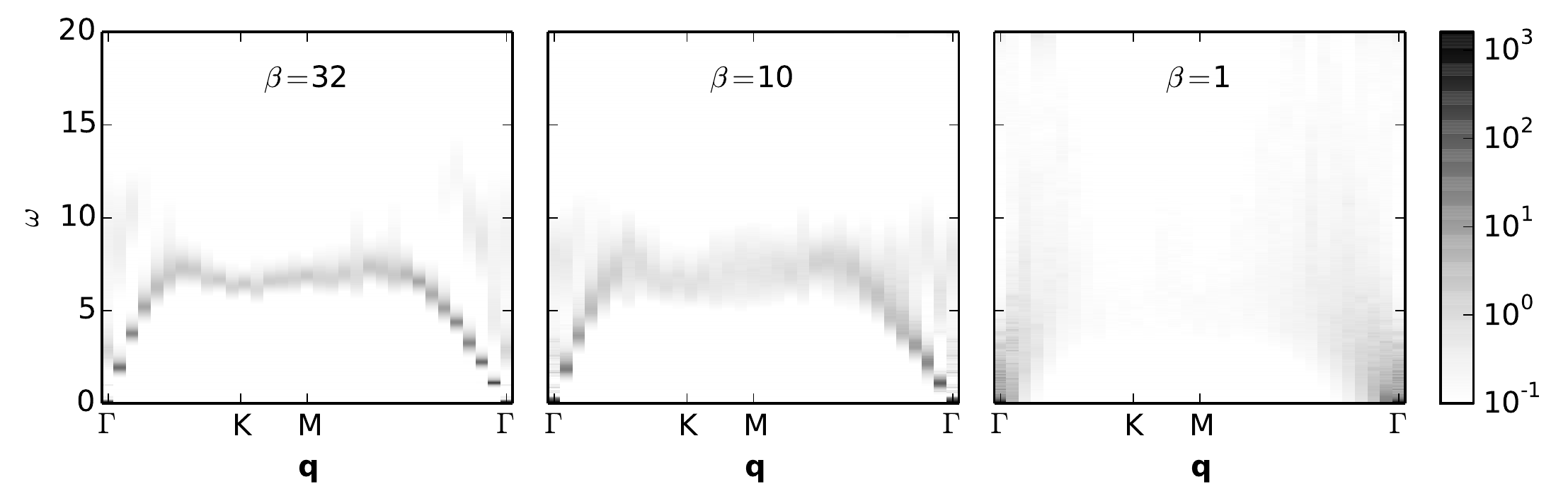}
  \caption{(Color online)  Dynamical quadrupolar structure factor $S_Q(\omega,\vect{q})$ at $\theta=-0.5\pi$ for different temperatures along the same Brillouin-zone path as in Fig.~5.}
  \label{fig:dsfT} 
\end{figure}

Within the quadrupolar phase, the integrated spectral weight $I_S(\vect{q})=\int S_S(\omega,\vect{q}) \: d\omega/(2\pi) $ vanishes linearly with $|\vect{q}|$ near the $\Gamma$-point, in accord with the flavor-wave theory result~\cite{laeuchli06}, as seen in  Fig.~\ref{fig:sfq}, where besides $I_S(\vect{q})$, we also show the corresponding quadrupolar quantity $I_Q(\vect{q})=\int S_Q(\omega,\vect{q}) \: d\omega/(2\pi) $ for several values of $\theta$. Here, we employed the fact that $I_X(\vect{q})=S_X(\vect{q})$ is conveniently obtained from the QMC equal-time correlations for both $X=S$ and $Q$. On the other hand, 
$I_Q(\vect{q})$ is seen to grow upon approaching the quadrupolar Bragg-peak position $\vect{q}=\Gamma=0$, where it diverges (cf. Sec.~III). The dispersing mode in $S_Q(\omega,\vect{q})$ thus provides direct access to the Goldstone mode in the excitation spectrum from the O(3) symmetry breaking in the quadrupolar phase. This low-energy quadrupolar wave exhibits a linear dispersion, and we estimate the corresponding velocity for the pure biquadratic model ($\theta=-\pi/2$) to be $c\approx 4.8$ 
from  our data shown in Fig.~\ref{fig:dsf}, a value that falls slightly below the value of $c=3\times 1.869(4)\approx 5.6$ estimated in Ref.~\onlinecite{kaul12} (after accounting for the factor $1/3$ in the value of $J$ actually employed in Ref.~\onlinecite{kaul12}), while flavor-wave theory gives a slightly smaller value of $c=\sqrt{18}\approx 4.2$ for $\theta=-\pi/2$. The  deviation from Ref.~\onlinecite{kaul12} is due to finite-size effects that restrict us from extracting the asymptotic value of $c$ from the observed dispersion near the $\Gamma$-point in Fig.~\ref{fig:dsf}, while the estimate in Ref.~\onlinecite{kaul12} was obtained from a finite-size analysis of winding number-based estimators. Within  flavor-wave theory as well as based on general Goldstone-mode counting considerations~\cite{nielsen76}, we conclude for the presence of \textit{two} such \textit{linear} Goldstone modes, and indeed, this count is in accord with the quantitative behavior of the low-$T$ specific heat, as discussed in Sec.~III. 
In Fig.~\ref{fig:dsfT}, we show the dynamical quadrupolar structure factor $S_Q(\omega,\vect{q})$ at $\theta=-0.5\pi$ for different temperatures. 
While the spectral weight within the higher energy range of the quadrupolar wave dispersion relation ($\omega \approx 5 - 10$) dissolves beyond 
$T\approx 0.1J$, one  still finds indications for low-energy soft-modes at temperatures  $T\approx J$, i.e. of order the exchange constant. This quantity thus provides a robust probe for  quadrupolar correlations in the system also at elevated temperature scales. 

Within the ferromagnetic regime, the low-energy Goldstone mode from the O(3) symmetry breaking is directly exhibited through the quadratically-dispersing soft-mode detected by $S_S(\omega,\vect{q})$, with a flat ($\vect{q}$-independent) integrated spectral weight $I_S(\vect{q})$, also shown in Fig.~\ref{fig:sfq}. Such a \textit{quadratic} magnon dispersion relation is again in accord with general considerations~\cite{nielsen76} as well as the $T$-linear low-temperature specific heat observed in the ferromagnetic regime (cf. Sec.~III).  

\begin{figure}[t]
  \centering
  \includegraphics[width=\columnwidth]{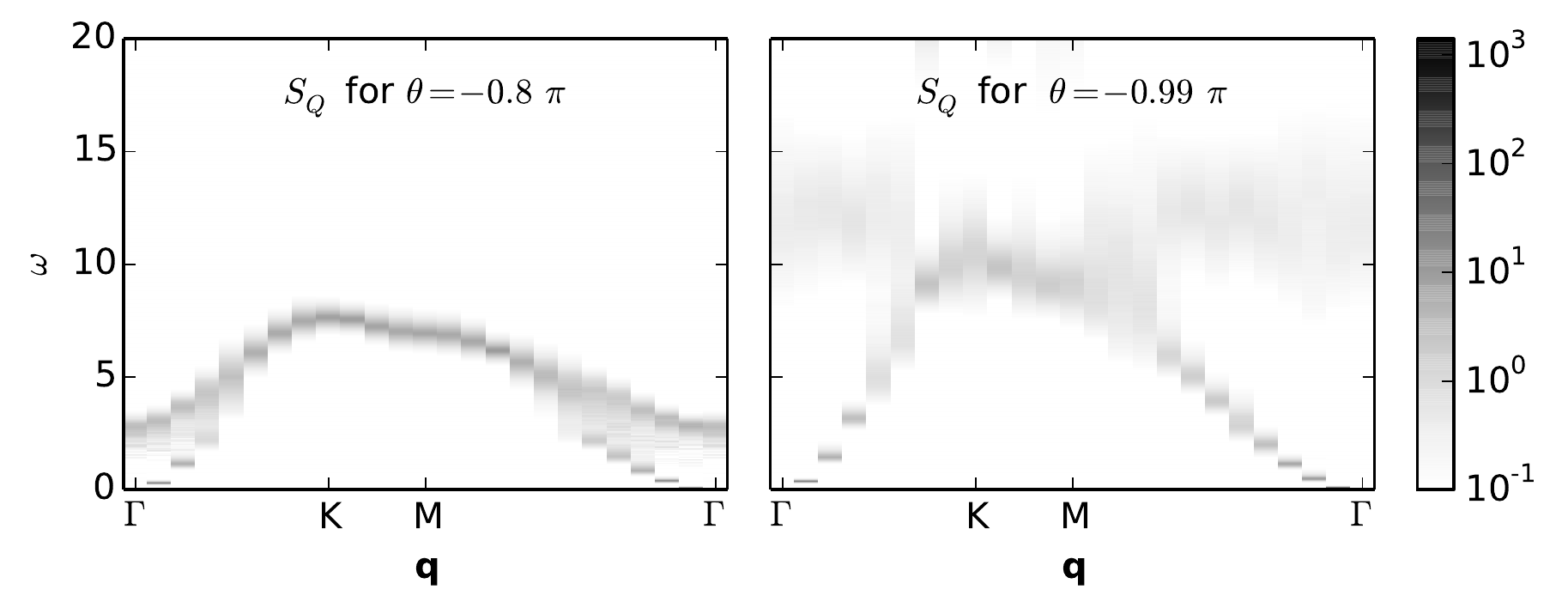}
  \caption{(Color online) Dynamical quadrupolar structure factor $S_Q(\omega,\vect{q})$ for different values of $\theta$ along the same Brillouin-zone path as in Fig.~5.}
  \label{fig:dsfmore} 
\end{figure}

In addition to the low-energy modes, that we discussed thus far, we also observe in $S_Q(\omega,\vect{q})$  additional, higher energy scattering weight, which is resolved most clearly when sufficiently separated from the low-energy mode, such as for $\theta<\theta_{SU(3)}$, within the ferromagnetic region. 
Additional spectral weight is also found at other values of $\theta$ than those shown in Fig.~\ref{fig:dsf}, cf. the data for $S_Q(\omega,\vect{q})$ shown in Fig.~\ref{fig:dsfmore}. 
Within the flavor-wave theory calculations of $S_Q(\omega,\vect{q})$ for the quadrupolar regime~\cite{pires14}, a continuum of scattering states from two-magnon contributions was obtained. In contrast to the QMC spectral functions, which exhibit a dominant spectral weight at low energies, the results in Ref.~\onlinecite{pires14} however exhibit also regions with a maximum in the spectral weight shifted towards the center of the continuum.   
On the other hand, our QMC spectral functions 
may be compared to the results in Ref.~\onlinecite{smerald13}. There, the model in Eq.~(\ref{hamiltonian}) was considered within the  antiferroquadrupolar regime, and calculations of the dynamical  quadrupolar  structure factor were performed both within 
quantum non-linear sigma-model as well as flavor-wave theory.
In addition to a dominant low-energy mode, a narrow, high-energy mode was observed in the dynamical quadrupolar  structure factor. 
Given the limited spectral resolution of the QMC spectral functions at elevated energies, our data would be consistent with a similar scenario also for the ferroquadrupolar parameter regime, accessed by our simulations. 

\section{Conclusions}
Employing a sign-problem free presentation of the quantum partition function, we performed quantum Monte Carlo simulations of the spin-1 bilinear-biquadratic Heisenberg model on the triangular lattice within a restricted parameter regime, which includes both ferromagnetic as well as ferroquadrupolar ordered regions. In addition, we examined the SU(3)-symmetric point that separates these regions. We identified the corresponding order parameters and extracted the leading low-temperature algebraic scaling of the specific heat, providing direct probes of to the Goldstone soft-modes. From quantum Monte Carlo calculations, combined with numerical analytic continuation, we obtained the dynamical spin and quadrupolar structure factors. These provided us with direct access to the  dispersion relations  of the Goldstone modes. Our numerical results are found to be in excellent agreement with the exact flavor-wave-theory results at the SU(3)-symmetric point. Furthermore,  we observed only weak quantitative deviations in the dispersion relations obtained within the ferroquadrupolar phase, such that  flavor-wave-theory is seen to provide an reasonably accurate  account of the low-energy modes. The dynamical quadrupolar structure factor was found to provide a useful probe for the ferroquadrupolar state also at elevated temperatures. In addition, we observed further spectral weight in the  dynamical quadrupolar structure factor, which is shifted towards high energies within the ferromagnetic regime. 
It will be interesting to study this distinct contribution to the spectral function by means of other approaches, which can access more accurately the high energy scattering weight.  
For the future, it will also be interesting to study possible continuous quantum phase transitions between spin nematic phases and paramagnetic or spin liquid regions in systems with competing interactions on non-bipartite lattices~\cite{nahum11,kaul12b}.

\acknowledgments

We acknowledge discussions with Ribhu K. Kaul, Andreas L\"auchli, Frederic Mila, and Karlo Penc, as well as support from the Deutsche Forschungsgemeinschaft within grants FOR 1807 and WE 3949/3-1. 
Furthermore, we acknowledge the allocation of CPU time through JARA-HPC at JSC J\"ulich and at RWTH Aachen University.

%
%

\begin{thebibliography}{99}


\bibitem{balents10} 
L. Balents, Nature (London) \textbf{464}, 199 (2010).


\bibitem{andreev84}
A. F. Andreev and I. A. Grishchuk, Zh. Eksp. Teor. Fiz. \textbf{87}, 467 (1984) [Sov. Phys. JETP \textbf{60}, 267 (1984)]


\bibitem{chubukov90}
A. Chubukov, J. Phys. Condens. Matter 2, 1593 (1990).


\bibitem{laeuchli05}
A. L\"auchli, J. C. Domenge, C. Lhuillier, P. Sindzingre, and M. Troyer, Phys. Rev. Lett. {\bf 95 }, 137206 (2005).


\bibitem{shannon06}
N. Shannon, T. Momoi, and P. Sindzingre, Phys. Rev. Lett. \textbf{96}, 027213 (2006).



\bibitem{mila00}
F. Mila and F. C. Zhang, Eur. Phys. J. B \textbf{16}, 7 (2000).


\bibitem{harada02}
K. Harada and N. Kawashima, Phys. Rev. B \textbf{65}, 052403 (2002).


\bibitem{bhattacharjee06}
S. Bhattacharjee, V. B. Shenoy, and T. Senthil, Phys. Rev. B \textbf{74}, 092406 (2006).


\bibitem{laeuchli06}
A. L\"auchli, F. Mila, and K. Penc, Phys. Rev. Lett. \textbf{97}, 087205 (2006).


\bibitem{tsunetsugu06}
H. Tsunetsugu and M. Arikawa, J. Phys. Soc. Jpn. \textbf{75}, 083701 (2006).


\bibitem{nakatsuji05}
S. Nakatuji, Y. Nambu, H. Tonomura, O. Sakai, S. Jonas,  C. Broholm, H. Tsunetsugu, Y. Qiu, Y. Maeno, Science Reports, {\bf 203}, 1697 (2005). 


\bibitem{nakatsuji10}
S. Nakatsuji, Y. Nambu, and S. Onoda, J. Phy. Soc. Jpn.{\bf 79}, 011003 (2010). 
 

\bibitem{stoudenmire09}
E. M. Stoudenmire, S. Trebst, and L. Balents, Phys. Rev. B \textbf{79}, 214436 (2009).


\bibitem{michaud11}
F. Michaud, F. Vernay, and F. Mila, Phys. Rev. B \textbf{84} 184424 (2011).


\bibitem{smerald13}
A. Smerald and N. Shannon, Phys. Rev. B {\bf 88}, 184430 (2013). 


\bibitem{smerald13b}
A. Smerald and N. Shannon, preprint arXiv:1303.4465 (2013), unpublished.



\bibitem{papanicolau84}
N. Papanicolaou, Nucl. Phys. B \textbf{240}, 281 (1984); 


\bibitem{joshi99}
A. Joshi, M. Ma, F. Mila, D. N. Shi, and F. C. Zhang, Phys. Rev. B \textbf{60}, 6584 (1999).

%


\bibitem{pires14}
A. S. T. Pires, Solid State Comm. {\bf 196}, 24 (2014). 


\bibitem{kawashima04}
N. Kawashima and K. Harada, J. Phys. Soc. Jpn. \textbf{73}, 1379 (2004).


\bibitem{harada01}
K. Harada and N. Kawashima, J. Phys. Soc. Jpn. \textbf{70}, 13 (2001).


\bibitem{harada07}
K. Harada, N. Kawashima, and M. Troyer, J. Phys. Soc. Jpn. \textbf{76}, 013703 (2007).


\bibitem{kaul12}
R. K. Kaul, Phys. Rev. B {\bf 86}, 104411 (2012).


\bibitem{syljuasen02}
O. F. Syljuasen and A. W. Sandvik, Phys. Rev. E \textbf{66}, 046701 (2002).


\bibitem{michel07}
F. Michel and H.G. Evertz, preprint arXiv:0705.0799 (2007), unpublished. 


\bibitem{beach04}
K. S. D. Beach, preprint arXiv:cond-mat/0403055 (2004), unpublished. 


\bibitem{mermin66}
N. D. Mermin and H. Wagner, Phys. Rev. Lett. {\bf 17}, 1133 (1966).


\bibitem{Kaulpc}
Ribhu K. Kaul, private communications.


\bibitem{Pencpc}
Karlo Penc, private communications.


\bibitem{nielsen76}
H. B. Nielsen and S. Chadha, Nucl. Phys. B {\bf 105}, 445 (1976). 


\bibitem{nahum11}
A. Nahum, J. T. Chalker, P. Serna, M. Ortu\~{n}o, and A. M. Somoza, Phys. Rev. Lett. {\bf 107}, 110601 (2011).


\bibitem{kaul12b}
R. K. Kaul, Phys. Rev. B {\bf 85}, 180411(R) (2012).


\end{thebibliography}
\end{document}